\documentclass[aps,prl,twocolumn,superscriptaddress,floatfix,amsmath,amssymb,nofootinbib,longbibliography]{revtex4-1}

\usepackage[utf8]{inputenc}
\usepackage[english]{babel}
\usepackage{amssymb,amsthm,amsmath,amstext,amsbsy,amsopn}
\usepackage{bbm}
\usepackage{graphicx}
\usepackage{isotope}

\usepackage{tabularx}
\newcolumntype{Y}{>{\centering\arraybackslash}X}

\usepackage{float}

\usepackage[dvipsnames]{xcolor}

\usepackage{hyperref}
\hypersetup{
    colorlinks = true,
    citecolor  = blue,
    linkcolor  = blue,
    urlcolor  = blue
}

\definecolor{plot1}{rgb}{0.86, 0.08, 0.24}
\definecolor{plot2}{rgb}{0.25, 0.41, 0.88}
\definecolor{plot3}{rgb}{1.0, 0.55, 0}
\definecolor{plot4}{RGB}{61,153,86}

\newcommand{\la}{\langle}
\newcommand{\ra}{\rangle}
\newcommand{\Xmax}[1]{\ensuremath{#1_\text{max}}}

\newcommand{\RSVD}{\ensuremath{R_\text{SVD}}}
\newcommand{\RSVDpercent}{\ensuremath{R^{\%}_\text{SVD}}}

\newcommand{\MeV}{\ensuremath{\text{MeV}}}

\newcommand{\fm}{\ensuremath{\text{fm}}}
\newcommand{\fmi}{\ensuremath{\text{fm}^{-1}}}

\newcommand{\elem}[2]{\ensuremath{^{#1}\text{#2}}}

\newcommand*{\nxlo}[1]{N${}^{#1}$LO}

\newcommand{\compression}[2]{\ensuremath{C^{#1}_{#2}}}

\newcommand{\ai}{\textit{ab initio}}
\newcommand{\ie}{\textit{i.e.}}
\newcommand{\eg}{\textit{e.g.}}

\newcommand{\JPT}{\mathcal{J}^\Pi\mathcal{T}}

\newcommand{\LamNNN}{\Lambda_\text{3N}}
\newcommand{\NN}{\text{NN}}
\newcommand{\NNN}{\text{3N}}

\begin{document}

\allowdisplaybreaks

\title{Randomized Low-Rank Decompositions of Nuclear Three-Body Interactions}

\author{A.~Tichai}
\email{alexander.tichai@physik.tu-darmstadt.de} 
\affiliation{Technische Universit\"at Darmstadt, Department of Physics, 64289 Darmstadt, Germany}
\affiliation{ExtreMe Matter Institute EMMI, GSI Helmholtzzentrum f\"ur Schwerionenforschung GmbH, 64291 Darmstadt, Germany}
\affiliation{Max-Planck-Institut f\"ur Kernphysik, Saupfercheckweg 1, 69117 Heidelberg, Germany}

\author{P.~Arthuis}
\email{parthuis@theorie.ikp.physik.tu-darmstadt.de}
\affiliation{Technische Universit\"at Darmstadt, Department of Physics, 64289 Darmstadt, Germany}
\affiliation{ExtreMe Matter Institute EMMI, GSI Helmholtzzentrum f\"ur Schwerionenforschung GmbH, 64291 Darmstadt, Germany}
\affiliation{Helmholtz Forschungsakademie Hessen für FAIR (HFHF), GSI Helmholtzzentrum für Schwerionenforschung GmbH, 64291 Darmstadt, Germany}

\author{K.~Hebeler}
\email{kai.hebeler@physik.tu-darmstadt.de}
\affiliation{Technische Universit\"at Darmstadt, Department of Physics, 64289 Darmstadt, Germany}
\affiliation{ExtreMe Matter Institute EMMI, GSI Helmholtzzentrum f\"ur Schwerionenforschung GmbH, 64291 Darmstadt, Germany}
\affiliation{Max-Planck-Institut f\"ur Kernphysik, Saupfercheckweg 1, 69117 Heidelberg, Germany}

\author{M.~Heinz}
\email{mheinz@theorie.ikp.physik.tu-darmstadt.de}
\affiliation{Technische Universit\"at Darmstadt, Department of Physics, 64289 Darmstadt, Germany}
\affiliation{ExtreMe Matter Institute EMMI, GSI Helmholtzzentrum f\"ur Schwerionenforschung GmbH, 64291 Darmstadt, Germany}
\affiliation{Max-Planck-Institut f\"ur Kernphysik, Saupfercheckweg 1, 69117 Heidelberg, Germany}

\author{J.~Hoppe}
\email{jhoppe@theorie.ikp.physik.tu-darmstadt.de}
\affiliation{Technische Universit\"at Darmstadt, Department of Physics, 64289 Darmstadt, Germany}
\affiliation{ExtreMe Matter Institute EMMI, GSI Helmholtzzentrum f\"ur Schwerionenforschung GmbH, 64291 Darmstadt, Germany}

\author{T.~Miyagi}
\email{miyagi@theorie.ikp.physik.tu-darmstadt.de}
\affiliation{Technische Universit\"at Darmstadt, Department of Physics, 64289 Darmstadt, Germany}
\affiliation{ExtreMe Matter Institute EMMI, GSI Helmholtzzentrum f\"ur Schwerionenforschung GmbH, 64291 Darmstadt, Germany}
\affiliation{Max-Planck-Institut f\"ur Kernphysik, Saupfercheckweg 1, 69117 Heidelberg, Germany}

\author{A.~Schwenk}
\email{schwenk@physik.tu-darmstadt.de}
\affiliation{Technische Universit\"at Darmstadt, Department of Physics, 64289 Darmstadt, Germany}
\affiliation{ExtreMe Matter Institute EMMI, GSI Helmholtzzentrum f\"ur Schwerionenforschung GmbH, 64291 Darmstadt, Germany}
\affiliation{Max-Planck-Institut f\"ur Kernphysik, Saupfercheckweg 1, 69117 Heidelberg, Germany}

\author{L.~Zurek}
\email{lzurek@theorie.ikp.physik.tu-darmstadt.de}
\affiliation{Technische Universit\"at Darmstadt, Department of Physics, 64289 Darmstadt, Germany}
\affiliation{ExtreMe Matter Institute EMMI, GSI Helmholtzzentrum f\"ur Schwerionenforschung GmbH, 64291 Darmstadt, Germany}

\begin{abstract}
First-principles simulations of many-fermion systems are commonly limited by the computational requirements of processing large data objects.
As a remedy, we propose the use of low-rank approximations of three-body interactions, which are the dominant such limitation in nuclear physics.
We introduce a novel randomized decomposition technique to handle the excessively large matrix dimensions and study the sensitivity of low-rank properties to interaction details.
The developed low-rank three-nucleon interactions are benchmarked in \ai{} simulations of few- and many-body systems.
Exploiting low-rank properties provides a promising route to extend the microscopic description of atomic nuclei to large systems where storage requirements exceed the computational capacities of the most advanced high-performance computing facilities.
\end{abstract}

\maketitle

\paragraph{Introduction.--} 
What is the most efficient way to formulate and solve the quantum many-body problem?
This question lies at the heart of many-body theory and impacts areas ranging from atomic physics, condensed matter physics, and quantum chemistry.
In nuclear physics, the \ai{} description of many-body systems has witnessed remarkable progress (see Refs.~\cite{Hebe15ARNPS,Herg20review,Hebe203NF}) such that nowadays up to hundred interacting particles and beyond can be targeted in first principles simulations~\cite{Morr17Tin,Miyagi2021,Arthuis2020a,Hu22Pb,Hebeler2022jacno,Tichai2023bcc}.
This great success is due to i) the development of high-precision nuclear interactions based on chiral effective field theory~\cite{Ente20Frontier,Epel19nuclfFront,Hebe203NF,Hebe11fits,Jian20N2LOGO}, and ii) the use of many-body expansion methods~\cite{Herg20review}. 
Basis expansions are widespread in many-body theory, the most common being many-body perturbation theory (MBPT)~\cite{Holt14Ca,Tich16HFMBPT,Tichai18BMBPT,Tichai2020review}, self-consistent Green's function theory~\cite{Dick04PPNP,Soma14GGF2N3N,Soma20SCGF,Barbieri2021}, coupled-cluster theory~\cite{Hage14RPP,Bind14CCheavy,Novario2020a,Tichai2023bcc}, and the in-medium
similarity renormalization group (IMSRG) approach~\cite{Herg16PR,Stroberg2019,Stroberg2021,Heinz2021}.
All these many-body frameworks have the great benefit of polynomial computational scaling with system size, thus circumventing the computational limitations of variational frameworks such as large-scale diagonalizations from configuration interaction approaches~\cite{Barr13PPNP} or real-space methods such as Quantum Monte Carlo~\cite{Carl15RMP}.

The fundamental input of basis expansion methods are matrix elements of second-quantized many-body operators and their computational complexity is directly linked to the dimensions of the data arrays that store the matrix elements.
Memory requirements quickly become prohibitive when the size of the single-particle basis is increased, \eg{}, in large systems and/or when symmetries of many-body operators are spontaneously broken~\cite{Novario2020a,Frosini2021mrI,Frosini2021mrII,Frosini2021mrIII,Hagen2022PCC,Yuan2022}. In particular the handling of three-body operators poses a significant computational challenge.
Such higher-mode tensors naturally emerge in the description of nuclear systems and cold atoms (see Refs.~\cite{Hamm13RMP,Hebe15ARNPS}) or as operators in high-precision calculations~\cite{Bind14CCheavy,Heinz2021}.
Therefore, the treatment of prohibitively large data objects is a universal problem in quantum many-body theory.

In other areas of quantum many-body research it has been recognized for a long time that the complexity of the many-body problem can be tamed by optimizing the representation of the underlying data objects by performing a so-called tensor decomposition, \ie{}, a rewriting of complex high-dimensional objects as a sum/product of objects with lower dimension~\cite{Kolda2009}.
Factorizations are at the heart of tensor-network theories like the density matrix renormalization group that fundamentally build upon a factorized wave-function ansatz (see Refs.~\cite{White1992,Schollwoeck2011,Baiardi2020}) and have been successfully applied in nuclear structure calculations~\cite{Papenbrock2005,Legeza2015,Foss17TetraN,Tichai2022dmrg,Tichai2024dmrg}.
In quantum chemistry, correlation expansions have been revisited in the context of factorized wave-function amplitudes (see Refs.~\cite{Kinoshita2003,Hohenstein2013,Schutski2017,Parrish2019,Lesiuk2021}) and tensor-decomposed electron-repulsion integrals have been used in electronic-structure calculations~\cite{Benedikt2011,Benedikt2013,Motta2019}.
While the design of factorized many-body frameworks poses a significant formal challenge, it provides a long-term perspective to problems in many-body theory.

In nuclear physics it was shown that modern nucleon-nucleon (\NN{}) interactions admit for excellent low-rank approximations by employing a singular-value decomposition (SVD) on the momentum-space matrix elements~\cite{Tich21SVDNN,Zhu21SVD}.
Such low-rank interactions provide an accurate description of nuclear observables in various applications ranging from two-nucleon phase shifts to infinite nuclear matter calculations and ground-state observables in medium-mass nuclei~\cite{Tich21SVDNN}.
Moreover, the two-body (Lippmann-Schwinger) scattering equations  were recently reformulated such that they can be solved in terms of the decomposition factors themselves without reconstructing the original matrix~\cite{Tichai2022svdlse}.
In a complementary way, more complex tensor decompositions have been explored to factorize nuclear many-body states~\cite{Tich19THC,Tichai2019pre}, and random embeddings of nuclear two-body operators have been used to sample perturbative estimates of the correlation energies of closed-shell nuclei~\cite{Zare2022jle}.

\begin{figure}[t!]
\includegraphics[width=.9\columnwidth]{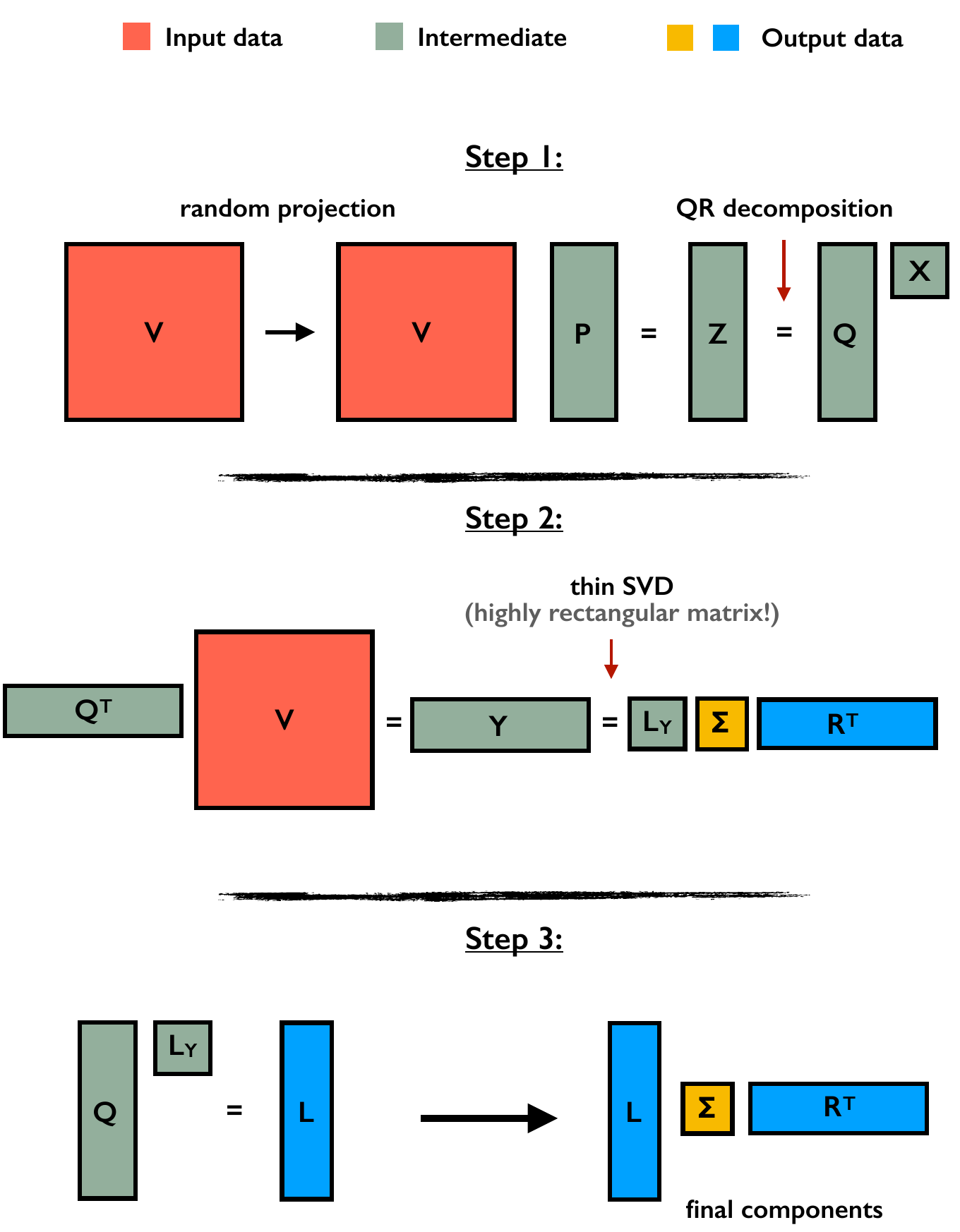}
\caption{Algorithmic steps in the calculation of a randomized SVD: 1) random projection and QR decomposition, 2) thin SVD of the projected interaction, 3) final evaluation of the left singular vectors.}
\label{fig:rsvd-scheme}
\end{figure}

In this work, we explore a scalable randomized linear-algebra solver that overcomes the limitations of traditional decomposition algorithms in terms of matrix dimension.
Using this novel method, low-rank approximations of three-nucleon (3N) forces are derived and benchmarked in few- and many-body applications.
Our results thus provide alternative, efficient storage formats of many-body operators and novel algorithms to obtain them.
While this Letter focuses on nuclei, the developments are discipline-agnostic and can be applied in other domains of many-body research.

\paragraph{Randomized matrix decompositions.--}
In the following we employ the singular value decomposition (SVD) of the 3N potential (with a dense matrix representation $V_\NNN \in \mathbb{R}^{N\times N}$ in a basis of size $N$)
\begin{align}
    V_\NNN = L \Sigma  R^T \, ,
\end{align}
where $L\, (R)$ denotes the unitary left (right) matrix of singular vectors and $\Sigma$ is a diagonal matrix containing the non-negative singular values $s_i$ ordered in decreasing size.

The conventional factorization of large matrices is a computationally intensive task. In this work we lower the computational cost of the decomposition by introducing a probabilistic component to the algorithm as proposed by Halko \textit{et al.}\ (see Ref.~\cite{Halko2011}).
The corresponding randomized SVD (rSVD) algorithm allows one to target the largest singular values much more efficiently.

A schematic representation of the rSVD algorithm is displayed in Fig.~\ref{fig:rsvd-scheme}.
In the first step a random projection matrix $P \in \mathbb{R}^{N\times l}$ is applied to the initial matrix $V$ yielding a matrix $Z=VP$ of much smaller size than $V$ -- in particular for low-rank matrices with $l\ll N$. 
Subsequently, an orthonormal basis of $Z$ is obtained by performing a QR decomposition $Z=QX$ (step 1).
The initial matrix $V$ is then projected to a smaller space using the low-rank basis $Q$ yielding $Y=Q^T V$.
Performing a thin SVD of the projected matrix yields $Y = L_Y \Sigma R^T$ (step 2). 
The left SVD component $L$ of the full decomposition is obtained by final left multiplication with $Q$, \ie{}, $L= Q L_Y$ (step 3).

In practice the number $l$ is chosen slightly larger than the desired rank $\RSVD$. The difference $p = l- \RSVD$ is termed oversampling parameter and is introduced to stabilize the computation and lower the sensitivity to the initial random sample. In our applications a value of $p=5$ is chosen.
The random character only enters through the choice of the initial projection operator $P$ and only a single run is performed in practice. While the repeated evaluation of the rSVD gives slightly different singular spectra, we validated our findings against an exact SVD for small matrices and found the random error to be negligible.
Our rSVD implementation is based on the \texttt{Eigen} library~\cite{eigenweb} such that all decompositions can be performed efficiently on a single compute node in a few hours.
A numerical implementation will be made publicly available in the future.

Once a decomposition has been obtained, the computational gain of the decomposition is quantified in terms of the compression ratio
\begin{align}
    \compression{N}{\RSVD} \equiv \frac{N^2}{2 N \RSVD + \RSVD} \, ,
\end{align}
where $N$ denotes the matrix dimension and $\RSVD$ the SVD-rank of the decomposition, \ie{}, the number of singular values kept.
The ratio \compression{N}{\RSVD} quantifies the memory savings of a rank-$\RSVD$ approximation over its original full-rank starting point of a square matrix with linear dimension $N$.
Since in the following we compare decompositions between matrices of different dimensions, it is useful to introduce the relative SVD-rank $\RSVDpercent \equiv 100 \cdot \RSVD/N$.
This is closely related to the compression ratio such that $\RSVDpercent=1.0$ corresponds to $C\approx100$ and $\RSVDpercent=0.1$ to $C\approx1000$.

\begin{figure*}[t!]
    \centering
    \includegraphics[width=1.\textwidth]{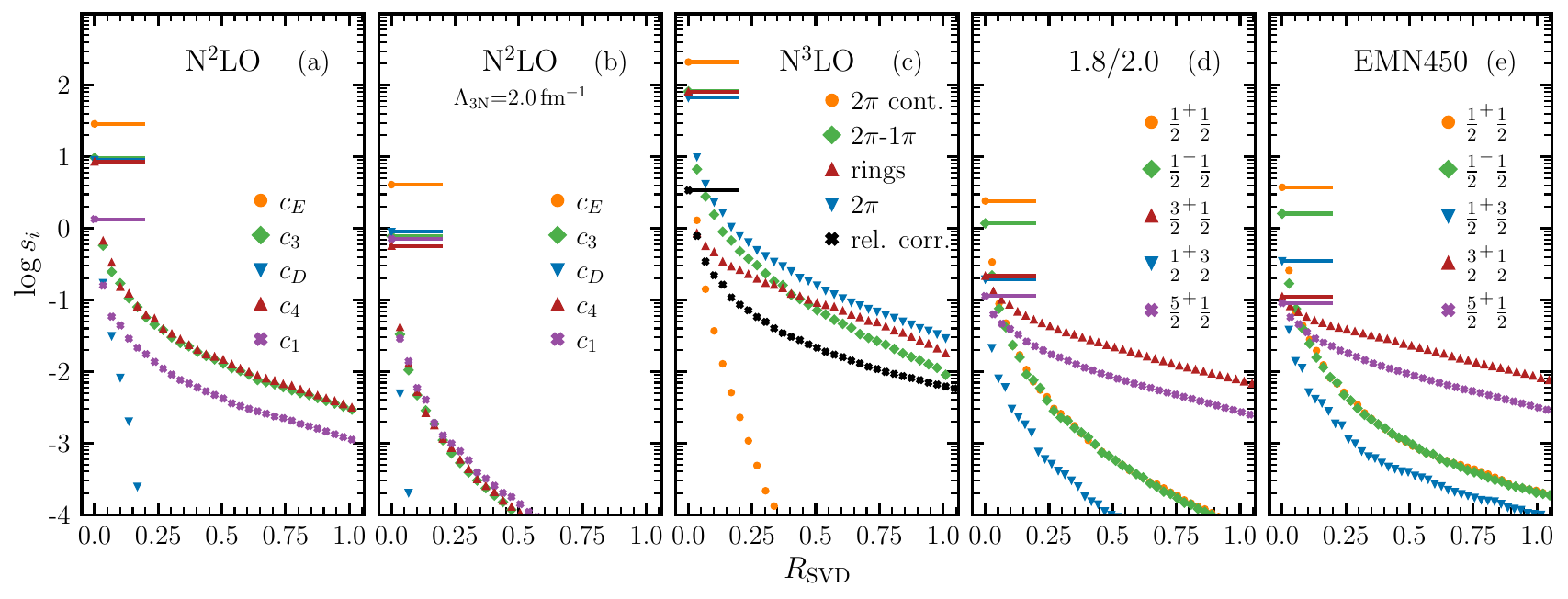}
    \caption{Logarithm of singular values as a function of truncation rank for different chiral 3N matrix elements: (a) unregularized \nxlo{2} operator topologies for the triton channel with all LECs set to one; (b) same as (a) but with a regulator of $\Lambda_{\NNN} = 2.0\, \fmi$; (c) unregularized \nxlo{3} topologies again with LECs set to one; and different partial-wave channels of (d) the ``1.8/2.0'' interaction and (e) the ``EMN450'' interaction (both including regulators). The leading singular value is indicated by a horizontal line. The legends are ordered according to the size of the largest singular value.}
    \label{fig:svals}
\end{figure*}

\paragraph{Low-rank interactions.--} In practice the rSVD is applied to the partial-wave-decomposed momentum-space matrix elements of 3N interactions
\begin{align}
    \la pq; \alpha | V_{\NNN} | p'q'; \alpha' \ra \, ,
    \label{eq:v3n}
\end{align}
where $p,q$ ($p',q'$) denote the three-body bra (ket) Jacobi momenta and the collective index $\alpha = \{ [(LS)J (ls)j]\mathcal{J} M_\mathcal{J} (Tt)\mathcal{T}M_\mathcal{T} \}$ gathers all quantum numbers of the three-body state.
Since the three-body matrix elements depend on four Jacobi momenta, bra and ket states are organized into collective indices $I=(pq\alpha)$ and $I'=(p'q'\alpha')$ such that a matrix-like object $M_{II'}$ is obtained from the initial $M_{pqp'q'}^{\alpha\alpha'}$.

For the partial-wave decomposition, angular-momentum eigenstates are obtained by coupling angular momenta of the first two particles to a two-body subsystem with quantum numbers $L$, $S$, $J$, which are subsequently coupled with the angular momenta of the third particle to obtain the three-body quantum numbers $\mathcal{J}$ and $\mathcal{T}$.
We emphasize that by working in a symmetry-adapted basis, we maintain rotational invariance and parity/isospin conservation in the factorization, \ie{}, $[\tilde V_{\NNN}, \mathcal{J}^2] = [\tilde V_{\NNN}, \mathcal{J}_z] =[\tilde V_{\NNN}, \Pi] =  [\tilde V_{\NNN}, \mathcal{T}^2] =  [\tilde V_{\NNN}, \mathcal{T}_z] = 0 \,$.
We note that a basis transformation mediated by a unitary matrix $U$ can be very efficiently performed starting from an initial SVD format, since the matrix multiplication $\bar V_{\NNN} \equiv U^\dagger V_{\NNN} U$ can be executed by operating only on the (thin) $L$ and $R^\dagger$ components of $V_{\NNN}$. This can enable speedups in such transformations, \eg{}, for generating single-particle matrix elements used in medium-mass calculations.

In the following, the partial-wave 3N matrix elements from Eq.~\eqref{eq:v3n} are taken as input for an rSVD solver.
For the discretization of the momentum mesh we use $N_p = N_q = 15$ points for both Jacobi momenta.
Intermediate two-body quantum numbers are included up to $J_\text{max}=8$.
Partial waves are characterized by their three-body quantum numbers $\JPT$.

Figure~\ref{fig:svals} displays the singular values of chiral 3N interactions.
Panel (a) shows the individual unregularized \nxlo{2} operator topologies in the triton channel, \ie{}, $\JPT=\frac{1}{2}^+\frac{1}{2}$, with the corresponding low-energy constants (LECs) set to unity in their respective natural units (see Ref.~\cite{Hebe203NF}).
We observe a clear difference between the set of long-range pion-exchange topologies $\{c_1,c_3,c_4\}$ that have a much slower falloff compared to the set $\{c_D,c_E\}$ that encode mid- and short-range contributions.
Still all topologies exhibit a strong suppression at very low ranks ($\lesssim 0.25 \, \%$) followed by a transition to a declining plateau-like region.
Even for the long-range topologies the size of the singular values is suppressed to less than $s_i \approx2\times 10^{-3}$ for the smallest $99 \%$. 
In particular in the case of the $c_E$ term all singular values are smaller than $10^{-15}$ except the largest one which is of the order ten.

In panel (b) we include non-local regulators $f_{\LamNNN}(p,q) = \exp[ - ((p^2 + \frac{3}{4} q^2)/\LamNNN)^4]$, where $\LamNNN$ denotes the three-body cutoff here set to $\LamNNN=2.0\, \fmi$. The regularized matrix elements induce a stronger suppression of the singular value spectrum for all operator topologies.
Similar findings as for \nxlo{2} are found at subleading orders [panel (c) for \nxlo{3}]: short-range relativistic corrections are more strongly suppressed than the long-range multi-pion exchange topologies. In general the \nxlo{3} terms have larger leading singular values and slower falloff in most cases.
Finally we show realistic chiral 3N forces (obtained after multiplication with the corresponding LEC values) in panel (d) for the 3N part of the ``1.8/2.0" interaction~\cite{Hebe11fits} with $\LamNNN = 2.0\, \fm^{-1}$ at \nxlo{2} and in panel (e) for the 3N part of the ``EMN450" interaction~\cite{Dris17MCshort} with $\LamNNN = 450\,$MeV at \nxlo{3}.

In general the regularization yields a stronger suppression of singular values, while this effect is more pronounced for smaller cutoff $\LamNNN$. 
Furthermore, a comparison with the individual operator topologies in panels (a) and (c) indicates that the singular spectrum is dictated by the dominating topology.
The slow falloff of the \nxlo{3} two-pion-exchange contributions yields a slower suppression of singular values in the case of the ``EMN450'' interaction.
For the full interactions we additionally investigate different partial-wave channels by varying angular momentum, parity, and isospin.
We note that the matrix dimensions among partial-wave channels differ significantly ($N=7650$ for $\JPT=\frac{1}{2}^+ \frac{3}{2}$ up to $N=40050$ for $\JPT = \frac{5}{2}^+\frac{1}{2}$).
As a general trend, higher partial-wave channels with larger $\mathcal{J}$ values show a slower falloff while their largest singular value is significantly smaller compared to, \eg{}, the triton channel.
To gauge the low-rank character of higher partial-wave channels we later explore many-body calculations.

\begin{figure}[t!]
    \centering
    \includegraphics[width=\columnwidth]{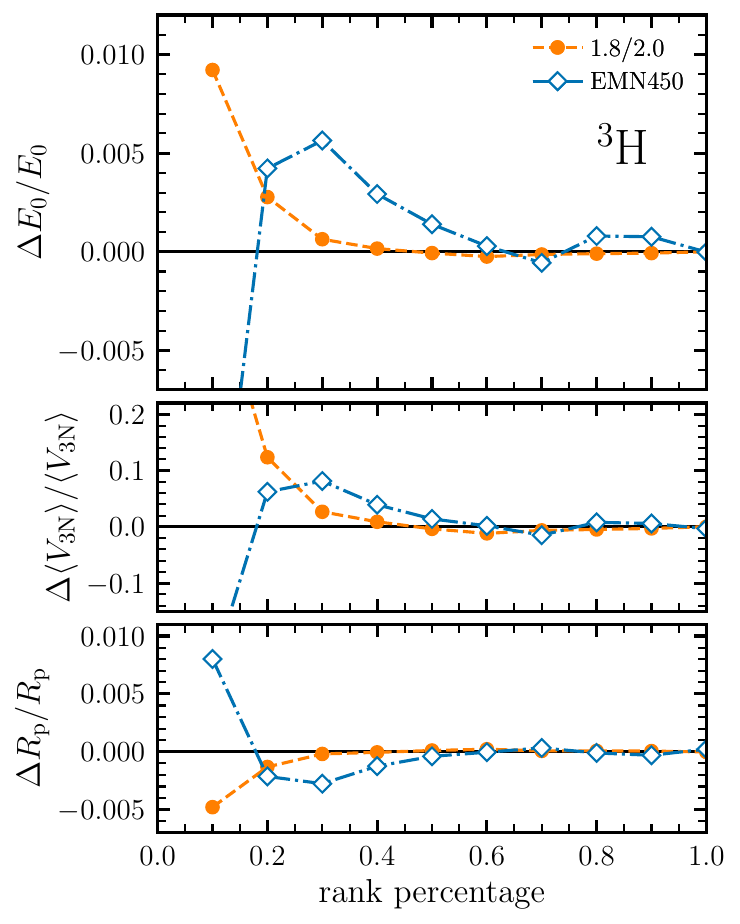}
    \caption{Relative error on the ground-state energy (top), expectation value of the 3N force (middle), and point-proton radius (bottom) for the triton as a function of SVD rank. The \NN{} interaction is not SVD-approximated. Results are shown for the ``1.8/2.0'' and ``EMN450'' interactions.}
    \label{fig:triton}
\end{figure}

\paragraph{Triton calculations.--}
A first benchmark for low-rank interactions is provided by the solution of the three-body problem for the ground-state energy and point-proton radius of the triton, using a Faddeev code in partial-wave-decomposed form~\cite{Stadler1991faddeev}.
Figure~\ref{fig:triton} provides an overview of the relative error on the ground-state energy (top), expectation value of the \NNN{} force (middle), and point-proton radius (bottom) as a function of decomposition rank for the ``1.8/2.0'' and ``EMN450'' interactions.
Similar to Fig.~\ref{fig:svals} we study the impact on the final observables by keeping up to $1\,\%$ of the leading singular values in the decomposition.
In the Faddeev calculation, we incorporate the full-rank \NN{} potential since we want to study the sensitivity to the low-rank properties of the \NNN{} force.
Since both interactions have been fitted to the triton binding energy, our full-rank calculations reproduce the experimental value of $E_\text{gs}(\elem{3}{H})$.
However, the individual size of the \NNN{} contribution is different among the interactions used, with $\la V_\NNN{} \ra_{1.8/2.0} = 0.19 \, \MeV$ and
$\la V_\NNN{} \ra_{\text{EMN}450} = 0.67\, \MeV$, respectively.
Consequently, we expect higher sensitivity in the case of the ``EMN450'' interaction in the following.

At low ranks ($\leq 0.5 \, \%$), the ``EMN450'' interaction yields consistently larger errors for all observables. This is in agreement with the slower falloff in the triton channel as observed in Fig.~\ref{fig:svals}.
Still, in all cases keeping the leading $1\, \%$ of singular values yields an excellent reproduction of ground-state properties with absolute differences of $10^{-4} \, \MeV$ ($\fm$) for the binding energy (point-proton radius).
In general the point-proton radius is less sensitive to the decomposition of the 3N forces since the 3N contribution is approximately $0.7 \% \,(6.4 \%)$ for the ``1.8/2.0'' (``EMN450'') interaction.
Few-body calculations thus allow for data compression of more than two orders of magnitude without significant loss in accuracy.

\begin{figure*}[t!]
    \centering
    \includegraphics[width=1\columnwidth]{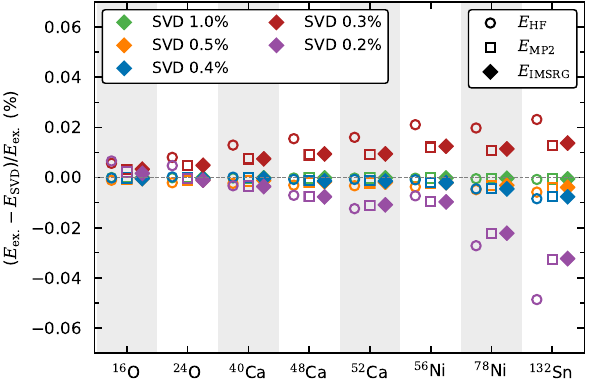} \quad
    \includegraphics[width=1\columnwidth]{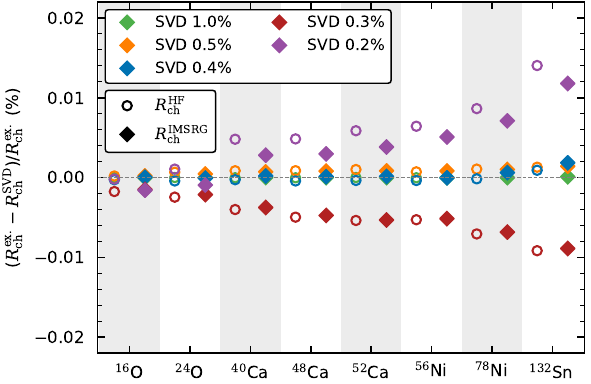}  
    \caption{Ground-state energies and charge radii at different many-body levels as a function of $\RSVDpercent{}$ for selected closed-shell nuclei using the ``1.8/2.0'' interaction. Calculations were performed in a large space with $\Xmax{e}=14$, $E_\text{3max}=24$, $\hbar \Omega=16 \, \MeV$. }
    \label{fig:midmass}
\end{figure*}

\paragraph{Medium-mass nuclei.--}
To test the performance of low-rank \NNN{} interactions for nuclei, we transform the matrix elements to a bound-state harmonic-oscillator basis.
Subsequently, we solve the Hartree-Fock equations with NN+3N interactions, thus obtaining an energetically optimized reference state. Using normal ordering, the resulting zero-, one-, and two-body parts of the intrinsic Hamiltonian define the starting point for an IMSRG calculation that non-perturbatively accounts for particle-hole correlations~\cite{Herg16PR}.
We employ the IMSRG(2) approximation, the current standard in IMSRG calculations, where operators are truncated at the normal-ordered two-body level.
Other observables are evaluated by a final transformation of the associated operators using the Magnus approach~\cite{Morr15Magnus}.
In addition, second-order MBPT results ($E_{\text{MP2}}$) are shown as a simpler estimate for the correlation energy.
Calculations are performed using the publicly available IMSRG solver by Stroberg~\cite{Stro17imsrggit}.

As a benchmark we compare ground-state energies and charge radii of selected closed-shell nuclei ranging from $^{16}$O to $^{132}$Sn with the developed low-rank 3N interactions.
All calculations employ 3N matrix elements with $E_{3\text{max}}=24$ using a storage format optimized for the normal-ordered two-body approximation~\cite{Miyagi2021}. Matrix element files are generated with the \texttt{NuHamil} code~\cite{Miyagi2023}.

Figure~\ref{fig:midmass} shows the relative error on the ground-state energy keeping $\RSVDpercent = 0.2, 0.3, 0.4, 0.5, 1.0$ of the leading singular values.
We observe in all cases that the impact of the SVD error is slightly larger at HF level compared to the MP2 and IMSRG(2) results.
For both ground-state energies and charge radii there is a systematic decrease of the relative error as a function of \RSVDpercent{}, although the dependence is non-monotonic.
Still, in all cases the relative error is well below $0.1\%$.
Moreover, a systematic increase of the relative error as a function of mass number is evident.
Since this trend is already present at the HF level, we attribute the increase in error to the enlarged hole-space in heavier systems and, thus, an enhanced accumulation of errors from individual matrix elements. The error on the correlation energy itself, both at the MP2 and IMSRG(2) level, does not increase in heavier systems.
Even for the heaviest systems, the SVD error is only at the few hundred keV level.

\paragraph{Summary and outlook.--} 
We explored low-rank matrix decompositions of three-body operators using a novel randomized SVD algorithm. This novel method is applied to chiral 3N interactions that are key in nuclear \textit{ab initio} calculations. We have clearly identified low-rank patterns in their momentum-space representation that enable efficient data compression.
In our few- and many-body benchmarks low-rank three-body interactions provide an excellent approximation for nuclear ground-state observables, thus paving the way for new efficient data representations of large many-body operators.
The memory savings that can be anticipated from our findings exceed a factor of $\compression{}{}=100$ at negligible approximation error on few- and many-body observables.

While the decomposition of many-body operators constitutes a natural first step, scaling benefits in actual applications require the reformulation of the many-body solver in terms of the decomposition factors themselves.
The design of a factorized Faddeev solver is a natural starting point to increase performance in few-body applications.
The transfer to the many-body sector constitutes a non-trivial open step that requires method-specific adaptations to provide computational gains. Factorization techniques define an innovative way of approaching new frontiers in nuclear physics~\cite{Heinz2021,Novario2020a}.
Our findings clearly motivate the transfer to other domains of many-body research; whenever low-rank properties of operators can be identified, factorization approaches may induce great scaling advantages. In that context the rSVD serves as a robust tool to find such patterns.

\begin{acknowledgments}
\paragraph*{Acknowledgements}
This work was supported in part by the European Research Council (ERC) under the European Union's Horizon 2020 research and innovation programme (Grant Agreement No.~101020842), the Deutsche  Forschungsgemeinschaft  (DFG, German Research Foundation) -- Projektnummer 279384907 -- SFB 1245, the Helmholtz Forschungsakademie Hessen für FAIR (HFHF) and by the BMBF Contract No.~05P21RDFNB.
The authors gratefully acknowledge the Gauss Centre for Supercomputing e.V.~(www.gauss-centre.eu) for funding this project by providing computing time through the John von Neumann Institute for Computing (NIC) on the GCS Supercomputer JUWELS at J\"ulich Supercomputing Centre (JSC).
\end{acknowledgments}

\bibliography{strongint}
\end{document}